\documentclass[pra,10pt,twocolumn,superscriptaddress,showpacs]{revtex4-1}

\usepackage{epsfig}
\usepackage{amsmath}
\usepackage{graphicx}
\usepackage{graphics}
\usepackage{float}
\usepackage{amssymb}
\usepackage[usenames,dvipsnames]{color}
\usepackage{natbib}
\usepackage{epstopdf}
\usepackage{verbatim}
\usepackage{wrapfig}

\newcommand{\bra}[1]{\left\langle #1\right|}
\newcommand{\ket}[1]{\left| #1\right\rangle}

\newcommand{\opav}[3]{\left\langle #1 | #2 | #3 \right\rangle}

\newcommand{\beq}{\begin{equation}}
\newcommand{\eeq}{\end{equation}}

\begin{document}

\title{Utilizing Nitrogen Vacancy Centers to measure oscillating magnetic fields}

\author{Adam Zaman Chaudhry}
\email{adamzaman@gmail.com}
\affiliation{Department of Electrical and Computer Engineering, National University of Singapore, 4 Engineering Drive 3, Singapore 117583}

\begin{abstract}
We show how nitrogen vacancy (NV) centers can be used to determine the amplitude, phase and frequency of unknown weak monochromatic and multichromatic oscillating magnetic fields using only the periodic dynamical decoupling (PDD) and Carr-Purcell-Meiboom-Gill (CPMG) sequences. The effect of decoherence on the measurement of the magnetic field parameters is explicitly analyzed, and we take into account the fact that different pulse sequences suppress decoherence to different extents. Since the sensitivity increases with increasing sensing time while it decreases due to decoherence, we use the Fisher information matrix in order to optimize the number of pulses that should be used.

\end{abstract}

\pacs{03.67.-a, 06.20.-f, 07.55.Ge, 85.75.Ss}
\date{\today}

\maketitle

\section{Introduction}

Measuring weak magnetic fields is an important problem with many applications in various fields such as data storage, biomedical sciences and material science \cite{FreemanScience2001}. For this task, nitrogen vacancy (NV) defect centers \cite{JelezkoApplPhys2002, DohertyPhysRep2013} have attracted considerable attention because they offer high magnetic field sensitivity due to their long coherence times, bio-compatibility and wide temperature range operation \cite{TaylorNatPhys2008, MazeNature2008, BalasubramanianNature2008, ChangNatNano2008, BalasubramanianNatNano2009, McguinnessNatNano2011, deLangePRL2011, HorowitzPNAS2012, HirosePRA2012, PhamPRB2012, NusranPRB2013, LoretzPRL2013, LeSageNature2013, GeiselmannNatureNano2013, MagesanPRA2013, CooperNatCommun2014, NusranarXiv2014}.

The basic idea behind the use of NV centers in magnetometry is very simple \cite{HallMRS2013, HongMRS2013, RondinarXiv2013}. Under suitable conditions, a NV center forms an effective two-level system. In the presence of a magnetic field, the energy difference between the two levels changes due to the Zeeman effect. If the NV center is prepared in a superposition of its two energy eigenstates, a phase, which is dependent on the magnetic field, develops between the energy levels. This phase difference can be converted into a population difference, which can be subsequently read out optically. Unfortunately, using NV centers in the manner just described is generally not possible. As a result of the interaction of the NV center with its surrounding environment, the superposition state undergoes decoherence, and the phase difference is lost too quickly. The coherence time of the NV center can be enhanced, however, by using dynamical decoupling techniques \cite{ViolaPRA1998,LloydPRL1999,BiercukNature2009,LiuNature2009}, whereby a sequence of rapid control pulses are applied to the NV center \cite{HansonScience2010,RyanPRL2010,NaydenovPRB2011,WanganddeLangePRB2012,BarGillNatCommun2012,ZhaoPRB2012,WitzelPRB2012}. These control pulses effectively remove the effect of the environment on the NV center. Unfortunately, they also remove the effect of a constant (or slowly varying) magnetic field. However, for rapidly oscillating fields the situation is different \cite{TaylorNatPhys2008}. If no control pulses are applied, the total phase averages out to zero since the magnetic field is oscillating - each time that the magnetic field reverses direction, in the Bloch sphere picture, the state of the NV center starts to rotate in the opposite direction. The action of each control pulse is also to reverse the direction of the rotating NV center state. The net result is that for an oscillating magnetic field and with control pulses applied, the phase difference accumulates. The important point is that now the effect of the environment is much smaller, thus leading to considerably longer coherence times, thereby enabling the use of NV centers for magnetometry.

To date, the focus has largely been on measuring oscillating fields of fixed frequency and known phase. High sensitivities can be achieved if the control pulses are tuned to the frequency of the magnetic field. However, it should be noted that we do not necessarily know both the amplitude \textit{and} the phase of the magnetic field. In fact, we need not even know the frequency, and indeed, the magnetic field can have a complicated temporal profile. As such, recently a scheme to construct the profile of arbitrary time-varying magnetic fields has been proposed \cite{CooperNatCommun2014,MagesanPRA2013}. Essentially, the idea is to apply many different control pulse sequences, associated with the Walsh functions \cite{Walsh1923}, with a fixed final acquisition time. The information obtained from each of these sequences is then used to reconstruct the magnetic field with excellent accuracy. On the other hand, the drawback is that a large number of different pulse sequences need to be utilized. Certainly, if the magnetic field is completely unknown, then a large number of different pulse sequences are required. However, if, for instance, we already have some information about the magnetic field, can we make do with very few pulse sequences? Reducing the resources required in the determination of the magnetic field is an important problem \cite{MagesanPRA2013b}. Consequently, in this paper, our objective is to measure oscillating fields of unknown amplitude, phase and frequency using only the bang-bang (BB) or periodic dynamical decoupling (PDD) sequence \cite{ViolaPRA1998} and the Carr-Purcell-Meiboom-Gill (CPMG) sequence \cite{CarrPhysRev1954,MeiboomRev1958}. Our contributions are the following. First, we derive expressions for the phase difference between the energy levels of a NV center due to a monochromatic magnetic field of arbitrary phase with the control pulses not necessarily tuned with the frequency of the magnetic field. Previously obtained expressions are valid for either control pulses tuned with the magnetic field and arbitrary phase or for pulses detuned from the magnetic field but zero phase for the magnetic field \cite{TaylorNatPhys2008,deLangePRL2011,NusranPRB2013}. We then show how using only these two pulse sequences, the amplitude, phase \textit{and} frequency of the field can be determined. Since our results are valid for arbitrary pulse intervals, they can be easily generalized to multichromatic fields. In particular, it is shown that $2M$ measurements, where $M$ is the number of frequency components in the field, are enough to determine the multichromatic magnetic field if the frequencies of the different components are known. We also show how to determine the frequencies in this case. By calculating the Fisher information, we show how the precision of the parameter estimates can be improved by increasing the number the pulses. However, all NV centers interact with their surrounding environment, and consequently decoherence affects our results. By computing the Fisher information matrix in the presence of decoherence using realistic parameters from recent experiments \cite{WanganddeLangePRB2012,MkhitaryanarXiv2014}, we are able to determine the optimal number of pulses that need to be applied for which the effects of increased sensing time and the degradation due to decoherence are balanced in order to obtain the best parameter estimates. 

This paper is organized as follows. In Sec.~II, we explain briefly how NV centers can be used as probes for detecting weak magnetic fields, and we describe the consequences of applying two different pulse sequences to the NV center. Next, in Sec.~III, we outline how the amplitude, phase and frequency of a monochromatic magnetic field could then be determined. In Sec.~IV, we generalize our results to multichromatic magnetic fields. The effects of decoherence on our proposed schemes is then analyzed in Sec.~IV. We conclude in Sec.~V. 

\section{The setup}

The negatively charged NV center consists of a vacancy defect and an adjacent substitutional nitrogen atom \cite{DohertyPhysRep2013}. The ground state of the NV center is a spin triplet, with the $m_S = \pm 1$ levels degenerate and the energy difference between the $m_S = 0$ and $m_S = \pm 1$ levels is equal to $2.87$ GHz. A static magnetic field along the quantization axis of the NV center can be applied to lift the degeneracy between the $m_S = \pm 1$ levels. We can then apply pulses which are in resonance between either the $m_S = 0$ and $m_S = 1$ sublevels or the $m_S = 0$ and $m_S = -1$ sublevels, thereby forming an effective two-level system (TLS). The interaction of this TLS with the magnetic field $B_z(t)$ that we want to measure can be described in terms of the angular momentum operator $S_z$ by the Hamiltonian 
\beq
\label{hamiltonian}
H(t) = 2\pi \gamma B_z(t) S_z,
\eeq
and $\gamma = 28 \, \text{Hz/nT}$ for the NV center. 
We consider the two relevant states to be $\ket{0}$ and $\ket{1}$ such that $\sigma_z \ket{0} = \ket{0}$ and $\sigma_z\ket{1} = -\ket{1}$, where $\sigma_z$ is the standard Pauli matrix.
As explained in the introduction, the idea is to prepare a superposition of states $\ket{0}$ and $\ket{1}$, evolve this state under the Hamiltonian \eqref{hamiltonian} so that a relative phase develops, and then to finally apply another pulse to convert this phase difference into an easily readable population difference. Essentially, this is Ramsey interferometry with the added caveat that, due to the oscillatory nature of the magnetic field and the interaction with the environment, it is essential to apply control pulses at intermediate times as well. We now discuss the application of the $\pi$ pulses more quantitatively. In particular, two different timing sequences shall be considered.

\subsection{Using $N$ PDD pulses}

The PDD or BB sequence corresponds to applying $[R(\pi)U(\tau)R(\pi)U(\tau)]^{N/2}$. This notation means that we evolve our state under the action of $U(\tau)$, which is the unitary time evolution operator corresponding to the Hamiltonian \eqref{hamiltonian}, for a time $\tau$, apply a $\pi$ pulse (corresponding to the unitary operator $e^{-i\pi \sigma_x/2}$), evolve again for time $\tau$, apply another $\pi$ pulse, and then repeat this whole cycle $N/2$ times. Accordingly, we have $N$ pulses in total. 

To begin, we apply a $\pi/2$ pulse (described by the rotation operator $e^{-i\pi\sigma_y/4}$) to the state $\ket{0}$. This prepares the coherent superposition 
$$\ket{\psi_i} = \frac{1}{\sqrt{2}}\left(\ket{0} + \ket{1}\right).$$
Now the unitary time-evolution operator corresponding to Eq.~\eqref{hamiltonian} for time evolution from time $t_1$ to time $t_2$ can be written as 
\begin{align}
\label{unitarytimeevolutionoperator}
U_0(t_2,t_1) &= \exp\left[ -i\int_{t_1}^{t_2} H(t') dt'\right] \notag \\
&= e^{-i\Phi[t_2,t_1]S_z},
\end{align}
with 
\begin{align}
\label{phasedifft2t1}
\Phi[t_2,t_1] &= \int_{t_1}^{t_2} 2\pi \gamma B_z(t') dt' \notag \\
&= -\frac{\gamma b}{f} \left[ \cos(2\pi f t_2 + \phi) - \cos(2 \pi f t_1 + \phi) \right],
\end{align}
where we have assumed that $B_z(t)$ is a simple monochromatic oscillating field, namely $B_z(t) = b \sin(2\pi ft + \phi)$. At this point, it is generally assumed that $\tau \rightarrow 1/2f$ (see, for instance, Ref.~\cite{NusranPRB2013}). In other words, the pulses are tuned to the frequency of the magnetic field. This can be motivated from the fact that for $\phi = 0$, the $\pi$ pulses are applied `in step' with changes in the magnetic field direction, leading to an accumulation of phase difference. Here, we will not be restricting ourselves to only $\tau \rightarrow 1/2f$. Rather, we will show how investigating other values of $\tau$ along with an arbitrary phase can help us not only in determining the frequency, but is also essential to accurately estimate multichromatic fields.

To determine the phase difference for general values of $\tau$, we first evolve the TLS for time $\tau$. Using Eq.~\eqref{unitarytimeevolutionoperator}, we find that the state $\ket{\psi_i}$ becomes (global phase factors, which do not have any physical consequence, are always discarded for simplicity)
$$ \frac{1}{\sqrt{2}} \left(\ket{0} + e^{i\Phi[\tau,0]}\ket{1} \right).$$
Applying the $\pi$ pulse has the effect of interchanging $\ket{0}$ and $\ket{1}$. Evolving for another time period $\tau$, we find that the quantum state is now
$$ \frac{1}{\sqrt{2}} \left( \ket{1} + e^{-i\Phi[2\tau,\tau]}e^{i\Phi[\tau,0]} \ket{0} \right).$$
After applying another $\pi$ pulse to obtain
$$ \frac{1}{\sqrt{2}} \left( \ket{0} + e^{-i\Phi(2\tau,\tau)}e^{i\Phi(\tau,0)} \ket{1} \right),$$
the first cycle is complete. It is then evident that the phase difference between states $\ket{0}$ and $\ket{1}$ is 
$$ \theta_1 = \Phi[\tau,0] - \Phi[2\tau,\tau].$$
Carrying on in the same way, we find that the phase difference after two cycles (or four pulses) is 
$$ \theta_2 = \Phi[\tau,0] - \Phi[2\tau,\tau] + \Phi[3\tau,2\tau] - \Phi[4\tau,3\tau]. $$
The pattern should now be clear. We can then immediately generalize to $N$ pulses,
\begin{align}
\theta_{N/2} = &\Phi[\tau,0] - \Phi[2\tau,\tau] + \hdots + \Phi[(N - 1)\tau,(N-2)\tau] \, - \notag \\
&\Phi[N\tau,(N-1)\tau].\notag
\end{align}
Using Eq.~\eqref{phasedifft2t1}, this can be written as 
\begin{align*}
\theta_{N/2} = &\frac{\gamma b}{f} \left[\cos(2\pi N f \tau + \phi) + \cos \phi \right] \, + \notag \\ 
&\frac{2\gamma b}{f} \sum_{k=1}^{N - 1} (-1)^k \cos(2\pi k f \tau + \phi).
\end{align*}
We now use the identity (see the Appendix)
\begin{align}
\label{firstidentity}
&\sum_{k = 1}^{N - 1} (-1)^k \cos(2\pi k f\tau + \phi) = \notag \\
&-\frac{1}{2} \sec(\pi f\tau)\left[ \cos(\pi f\tau + \phi) + \cos(2\pi N f\tau - \pi f\tau + \phi) \right] 
\end{align}
to obtain, after further simplification \footnote{Strictly speaking, Eq.~\eqref{thetaBB} becomes ill-defined when $\tau = 1/2f$ because for this case, in the derivation, we divide by zero. However, the limit $\tau \rightarrow 1/2f$ exists and gives the same answer as if we had set $\tau = 1/2f$ from the beginning. A similar reasoning holds for Eq.~\eqref{thetaCP}.},
\beq
\label{thetaBB}
\theta_{\text{BB}} \equiv \theta_{N/2} = \frac{\gamma b}{f} \tan(\pi f \tau) \left[ \sin\phi - \sin(2\pi N f\tau + \phi)\right].
\eeq
It should be noted that we do not consider the final time $T = N\tau$ to be fixed. In this regard, our methods are different from Refs.~\cite{CooperNatCommun2014,MagesanPRA2013}, and more in the spirit of Refs.~\cite{deLangePRL2011, WanganddeLangePRB2012}. We emphasize again that we have not assumed $\tau = 1/2f$, and the phase $\phi$ is arbitrary as well.

In order to read out this phase difference, we note that the quantum state at the end of $N/2$ cycles is 
$$ \ket{\psi_{N/2}} = \frac{1}{\sqrt{2}} \left( \ket{0} + e^{i\theta_{\text{BB}}}\ket{1}\right). $$
We then apply a $\pi/2$ pulse, given by $e^{-i\frac{\pi}{2}\sigma_x}$, and measure the observable $\sigma_z$, with eigenvalues $\pm 1$. For the probability of obtaining eigenvalue $n$, we can write
\begin{equation}
\label{probabilities}
p(n|\theta_{\text{BB}}) = \frac{1}{2}\left[ 1 + n \sin \theta_{\text{BB}} \right],
\end{equation}
and we find that 
\begin{align}
&\langle\psi_{N/2}|e^{i\pi \sigma_x /4} \sigma_z e^{-i \pi \sigma_x /4}|\psi_{N/2}\rangle = \opav{\psi_{N/2}}{\sigma_y}{\psi_{N/2}} \notag \\
&= \sin(\theta_{\text{BB}}),
\end{align}
which is approximately equal to $\theta_{\text{BB}}$ for weak magnetic fields. This means that there could be an ambiguity about the magnetic field as the phase difference is bounded \cite{NusranPRB2013,NusranarXiv2014}. 

From the probabilities given by Eq.~\eqref{probabilities}, we can estimate $b$, $\phi$ and $f$ since information about these parameters is encoded in $\theta_{\text{BB}}$. How well we can estimate these parameters, however, is given by the Fisher information matrix \cite{Bosbook}. Defining $\mathbf{y}$ to be the vector of parameters to be estimated and $l(\mathbf{y}) = \ln p(n|\mathbf{y})$, the Fisher information matrix is defined as 
\begin{equation}
\mathbf{I}(\mathbf{y}) = -E \left[ \frac{\partial^2 l(\mathbf{y})}{\partial \mathbf{y}^2} \right],
\end{equation}
where $E$ denotes taking the average with respect to $p(n|\mathbf{y})$. The Fisher information matrix is useful because it sets bounds on the precision with which the various parameters can be estimated - the greater the Fisher information, the more precise are our estimates. In particular, the Cramer-Rao bound tells us that for any unbiased estimate of $\mathbf{y}$, $\text{cov}(\mathbf{y}) \geq \mathbf{I}^{-1}(\mathbf{y})$. Using the form of $p(n|\mathbf{y})$ given in Eq.~\eqref{probabilities}, we find that 
\begin{equation}
\label{Iintermsofderivatives}
[I^{\text{BB}}(\mathbf{y})]_{mn} = \frac{\partial \theta_{\text{BB}}}{\partial y_m} \frac{\partial \theta_{\text{BB}}}{\partial y_n}.
\end{equation}
Note that we have explicitly considered a particular measurement, which is actually the measurement performed in experiments, in order to compute the Fisher information. An important question is: can we do any better by performing a different quantum measurement? To show that the answer is no, we compute the quantum Fisher information matrix, which is the Fisher information optimized over all the possible POVMs that can be performed. The quantum Fisher information matrix is given by \cite{Helstrombook,Holevobook,CavesPRL1994}
\begin{equation}
[I(\mathbf{y})]_{mn} = \text{Tr}[\partial_{y_m} \rho_{\mathbf{y}} L_{y_j}],
\end{equation}
where $\rho_{\mathbf{y}} = \sum_k \rho_k \ket{\psi_k}\bra{\psi_k}$ is the quantum state from which we are estimating the parameters $\mathbf{y}$, and 
\begin{equation}
L_{y_m} = 2\sum_{k,l} \frac{\opav{\psi_k}{\partial_{y_m}\rho_{\mathbf{y}}}{\psi_l}}{\rho_k + \rho_l} \ket{\psi_k}\bra{\psi_l}
\end{equation}
for $\rho_k + \rho_l \neq 0$. For our quantum state, we find that 
\begin{equation*}
\partial_{y_m}\rho_{\mathbf{y}} = \frac{1}{2} \left( \begin{array}{ccc}
0 & -i\frac{\partial \theta_{\text{BB}}}{\partial y_m} e^{-i \theta_{\text{BB}}} \\
i\frac{\partial \theta_{\text{BB}}}{\partial y_m} e^{i \theta_{\text{BB}}} & 0 \end{array} \right).
\end{equation*}
Diagonalizing $\rho_{\mathbf{y}}$, we find two eigenvalues $\rho_1 = 0$ and $\rho_2 = 1$ corresponding to the eigenvectors $\ket{\psi_1} = 1/\sqrt{2} [1 \; -e^{i\theta_{\text{BB}}}]^T$ and $\ket{\psi_2} = 1/\sqrt{2} [e^{-i\theta_{\text{BB}}} \; 1]^T$. This leads to 
\begin{equation*}
\opav{\psi_1}{\partial_{y_m}\rho_{\mathbf{y}}}{\psi_2} = -\frac{i}{2} e^{-i\theta_{\text{BB}}} \frac{\partial \theta_{\text{BB}}}{\partial y_m}.
\end{equation*}
The quantum Fisher information matrix is then found to be 
\begin{align}
[I(\mathbf{y})]_{mn} &= 2[\opav{\psi_1}{\partial_{y_n}\rho_{\mathbf{y}}}{\psi_2} \opav{\psi_2}{\partial_{y_m}\rho_{\mathbf{y}}}{\psi_1}  \notag \\
&+ \opav{\psi_2}{\partial_{y_n}\rho_{\mathbf{y}}}{\psi_1} \opav{\psi_1}{\partial_{y_m}\rho_{\mathbf{y}}}{\psi_2}] \notag \\
&=\frac{\partial \theta_{\text{BB}}}{\partial y_m} \frac{\partial \theta_{\text{BB}}}{\partial y_n},
\end{align} 
which is thus the same as the Fisher information matrix we calculated using the explicit measurement scheme described before.   

The Fisher information matrix is now explicitly calculated, but the detailed form for general $\tau$ is rather complicated. However, once we set $\tau$ to be close to $1/2f$ in the general expressions, great simplifications occur. Suppose that we are estimating both $b$ and $\phi$ for a monochromatic magnetic field. We write the matrix as 
\[ \left( \begin{array}{cc}
I^{\text{BB}}_{bb} & I^{\text{BB}}_{b\phi} \\
I^{\text{BB}}_{\phi b} & I^{\text{BB}}_{\phi \phi} \end{array} \right).\]
Since the matrix is symmetric, we only note that 
\begin{align}
I^{\text{BB}}_{bb} &= \frac{4N^2\gamma^2}{f^2} \cos^2 \phi, \notag \\
I^{\text{BB}}_{\phi \phi} &= \frac{4N^2 \gamma^2 b^2}{f^2} \sin^2 \phi, \notag \\
I^{\text{BB}}_{b\phi} &= - \frac{2N^2 \gamma^2 b}{f^2} \sin(2\phi).
\end{align} 
Suppose that we only want to estimate a single parameter $y_j$. Then we have, using the Cramer-Rao bound for any unbiased estimator of $y_j$,
\begin{equation}
\text{Var}(y_j) \geq \frac{1}{I_{jj}}.
\end{equation} 
For example, if we are estimating only the amplitude, then 
$$ \text{Var}(b) \geq \frac{f^2}{4N^2 \gamma^2 \cos^2 \phi}. $$
It is important to realize that if both $b$ and $\phi$ are unknown, then these parameters cannot be estimated using the BB sequence alone. This is reflected in the fact that for this case, the Fisher information matrix becomes singular. 
Also, we have currently not taken decoherence into account. As will see in Sec.~\ref{sectiondecoherence}, decoherence comes into play for large $N$. We then need to optimize the number of pulses that we are applying in order to maximize the Fisher information, and thus to obtain the best estimates.

\subsection{Using $N$ CPMG pulses}

The CPMG sequence, given by $[U(\tau/2)R(\pi)U(\tau)R(\pi)U(\tau/2)]^{N/2}$, looks quite similar to the previous BB sequence. They differ due to only two segments at the beginning and the end of the sequences. Yet this seemingly small difference leads to very different results for the phase difference, and for the suppression of decoherence. We now find that after one cycle, 
$$ \theta_1 = \Phi[\tau/2,0] - \Phi[3\tau/2,\tau/2] + \Phi[2\tau,3\tau/2]. $$
Going to $N/2$ cycles,  
\begin{align*}
&\theta_{N/2} = \Phi[\tau/2,0] - \Phi[3\tau/2,\tau/2] + \Phi[2\tau,3\tau/2] + \hdots + \\ \notag 
&\Phi[(N-3/2)\tau,(N-2)\tau]  +  \Phi[N\tau,(N-1/2)\tau] \, - \\ \notag
&\Phi[(N-1/2)\tau,(N-3/2)\tau].
\end{align*}
Using Eq.~\eqref{phasedifft2t1}, this can be written as 
\begin{align*}
\theta_{N/2} = &-\frac{\gamma b}{f} \left[\cos(2\pi N f \tau + \phi) - \cos \phi \right] \, - \notag \\ &\frac{2\gamma b}{f} \sum_{k=0}^{N - 1} (-1)^k \cos[\pi (2k+1) f \tau + \phi].
\end{align*}
We now use the identity 
\begin{align}
\label{secondidentity}
&\sum_{k = 0}^{N - 1} (-1)^k \cos[(2k + 1)\pi f\tau + \phi] = \frac{1}{2} \sec(\pi f\tau) \, \times \notag \\
&\left[ \cos\phi - \cos(2\pi N f\tau + \phi) \right], 
\end{align}
to obtain, after further simplification,
\begin{align}
\label{thetaCP}
\theta_{\text{CP}} \equiv \theta_{N/2} = &\frac{\gamma b}{f} \left[\sec(\pi f\tau) - 1\right] \, \times \notag \\
&\left[\cos(2\pi Nf\tau + \phi) - \cos \phi\right].
\end{align}
Once again, we can compute the Fisher information matrix for $\tau$ close to $1/2f$. We now have 
\begin{align}
I^{\text{CP}}_{bb} &= \frac{4N^2\gamma^2}{f^2} \sin^2 \phi, \notag \\
I^{\text{CP}}_{\phi \phi} &= \frac{4N^2 \gamma^2 b^2}{f^2} \cos^2 \phi, \notag \\
I^{\text{CP}}_{b\phi} &= \frac{2N^2 \gamma^2 b}{f^2} \sin(2\phi).
\end{align} 
As before, using the CPMG sequence allows us to estimate only one of the parameters with finite Cramer-Rao lower bound. 

\section{Determining the amplitude, phase and frequency of the field}

\subsection{Frequency known}
Let us start by assuming that we have some idea what the frequency of the field $f$ is. We will show later how the frequency can be determined. What we want to show here is that our expressions reduce to the well-known results for $\tau \rightarrow 1/2f$, following which the phase and the amplitude of the monochromatic field can be determined \cite{NusranPRB2013}. It is straightforward to show that for $\tau = \frac{1}{2f}\left( 1 + \Delta \right)$, in the limit $\Delta \rightarrow 0$, $\theta_{\text{BB}} = \frac{2N\gamma b}{f} \cos\phi$ and $\theta_{\text{CP}} = \frac{2N\gamma b}{f} \sin\phi$, which gives 
\beq
\tan\phi = \frac{\theta_{\text{CP}}(\tau \rightarrow 1/2f)}{\theta_{\text{BB}}(\tau \rightarrow 1/2f)}.
\eeq
Defining $\theta_{\text{BB,CP}} = \sqrt{\theta_{\text{BB}}^2 + \theta_{\text{CP}}^2}$, we have that  
\begin{equation}
b = \frac{\theta_{\text{BB,CP}}f}{2N\gamma}.
\end{equation}
To quantify how well we can estimate the parameters specifying the AC field, we can compute the total Fisher information matrix which uses the results from both the BB and CPMG sequences. This is obtained from 
\begin{equation}
\mathbf{I}^{\text{BB,CP}} = \mathbf{I}^{\text{BB}} + \mathbf{I}^{\text{CP}},
\end{equation}
leading to
\begin{align}
&I^{\text{BB,CP}}_{bb} = \frac{4N^2\gamma^2}{f^2}, \notag \\
&I^{\text{BB,CP}}_{\phi \phi} = \frac{4N^2 \gamma^2 b^2}{f^2}, \notag \\
&I^{\text{BB,CP}}_{b\phi} = 0.
\end{align} 
This matrix is non-singular, meaning that we can estimate both $b$ and $\phi$ with finite precision. Furthermore, $I^{\text{BB,CP}}_{bb}$ and $I^{\text{BB,CP}}_{\phi \phi}$ both do not depend on $\phi$. Using the Cramer-Rao bound, we find that for any unbiased estimators, 
\begin{align}
\text{Var}(b) &\geq \frac{f^2}{4N^2 \gamma^2}, \notag \\
\text{Var}(\phi) &\geq \frac{f^2}{4N^2 \gamma^2 b^2}.
\end{align}

\subsection{Frequency unknown}
If the frequency is not known, we first need to find some way of estimating the frequency. One possible way is to note that, for fixed $N$, $\theta_{\text{BB}}$ is a periodic function of $\tau$ with period $1/f$. Near $\tau = 1/2f$, $\theta_{\text{BB}}$ has a peak; it then follows that $\theta_{\text{BB}}$ also has a peak near $3/2f$. By measuring the distance between two peaks, it is then possible to find the frequency, and to then find out the phase and amplitude as explained above. This method assumes that we are able to increase the total sensing time $T = N\tau$ such that the coherence has not become negligible at time $T$.

\begin{figure}[t]
   \includegraphics[scale = 0.6]{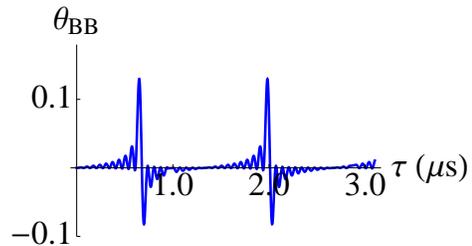}
   \centering
  	\caption{Plot of $\theta_{\text{BB}}$ against $\tau$. We use $f = 0.75 \text{MHz}$, $b = 0.1 \mu T$, and choose $\phi = \pi/3$. We have applied $N = 20$ pulses in this case according to the bang-bang pulse scheme, with the pulse interval given by $\tau$.}
  	\label{thetaBBphipiby3}
\end{figure}

\begin{figure}[t]
   \includegraphics[scale = 0.6]{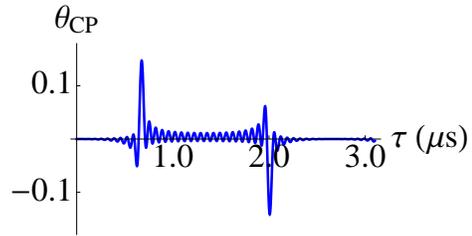}
   \centering
  	\caption{Plot of $\theta_{\text{CP}}$ against $\tau$. The parameters used are the same as Fig.~\ref{thetaBBphipiby3}. The first peak is located at $\tau \approx 0.676 \, \mu$s, while the second peak is located at $\tau \approx 2.008 \, \mu$s. The difference in peaks is not exactly equal to $1/f = 1.333 \mu$s.}
  	\label{thetaCPphipiby3}
\end{figure}

We illustrate this method in Fig.~\ref{thetaBBphipiby3}. Choosing $\phi = \pi/3$, we investigate the behaviour of $\theta_{\text{BB}}$ as a function of $\tau$. We find two peaks at approximately $\tau = 0.65 \,\mu$s and at $\tau = 1.98 \, \mu$s, which are near $\tau = 1/2f = 0.67 \,\mu$s and $\tau = 3/2f = 2 \,\mu$s respectively. However, the distance between the peaks is $1/f = 1.33 \,\mu$s, as expected. On the other hand, $\theta_{\text{CP}}$ does not share this periodic feature [see Fig.~\ref{thetaCPphipiby3}]. Nevertheless, the first peak is still located near $1/2f$. The precision of the estimate of frequency calculated in this manner can be found from calculating $I_{ff}^{\text{BB}}$ for $\tau \rightarrow 1/2f$ and $\tau \rightarrow 3/2f$, leading to
\begin{align}
I_{ff} &=  \frac{N^2 \gamma^2 b^2}{f^4} \, \times \notag \\
&(10N^2 \pi^2 \sin^2 \phi + 8N\pi \sin 2\phi + 4 \cos^2 \phi).
\end{align}
Once again, increasing $N$ increases the Fisher information, thereby leading to a more precise estimate of $f$.

Alternatively, since both $\theta_{\text{BB}}$ and $\theta_{\text{CP}}$ display a peak near $1/2f$, we can also obtain a good estimate of the frequency by finding out the behaviour of $\theta_{\text{BB,CP}} \equiv \sqrt{\theta_{\text{BB}}^2 + \theta_{\text{CP}}^2}$. This function shows a peak near $1/2f$ regardless of phase. Moreover, the position of this peak becomes closer and closer to $1/2f$ as $N$ is increased. More specifically, it can be shown by setting $\tau = \frac{1}{2f}(1 + \Delta)$, that for large $N$ the peak is located approximately at 
\beq 
\Delta \approx \frac{6 \sin^2 \phi}{\pi(N^2 - 6N \sin 2\phi + 3 \cos 2\phi - 1)}.
\eeq
Obviously, as $N$ increases, $\Delta$ becomes smaller and smaller, which means that the position of the peak almost coincides with $\tau = 1/2f$. Again, once the frequency is known, the phase and the amplitude can be figured out.

\begin{figure}[t]
   \includegraphics[scale = 0.6]{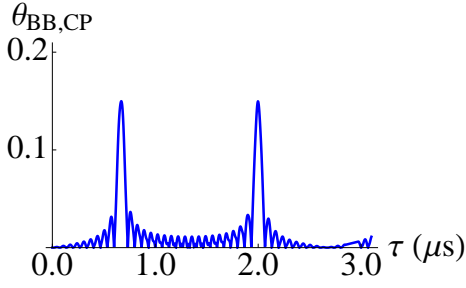}
   \centering
  	\caption{Plot of $\theta_{\text{BB,CP}}$ against $\tau$. We have $f = 0.75 \text{MHz}$, $b = 0.1 \, \mu\text{T}$ and choose $\phi = \pi/3$. We have applied $N = 20$ pulses.}
  	\label{thetaBBCPphipiby3}
\end{figure}

In Fig.~\ref{thetaBBCPphipiby3}, we have plotted $\theta_{\text{BB,CP}}$ as a function of $\tau$. for $N = 20$. The actual frequency of the AC field is $f = 750$ kHz. The location of the peak is approximately at $\tau = 0.6697 \, \mu$s. This gives us $f \approx 747$ kHz, which is an error of approximately $0.4\%$. Increasing $N$ further improves the measurement of $f$. However, $N$ should not be increased so much that $N \tau$ exceeds the coherence time of the NV center. Also, we now find that the Fisher information is
\begin{align*}
I_{ff} &= \left( \frac{N\gamma b}{f^2}\right)^2 \, \times \notag \\
&[\pi^2(N^2 + N\sin 2\phi + \sin^2\phi) + 4\pi \sin^2\phi + 4].
\end{align*}

\section{Beyond Monochromatic AC fields}
We now consider magnetic fields of the form 
$$ B_z(t) = \sum_{m=1}^M b_m \sin(2\pi f_m t + \phi_m). $$
It should be noted that due to the different frequency components of the magnetic field, we can in general apply control pulses that are tuned with only one of the frequencies. Consequently, we must explicitly consider the contribution of the magnetic field that is detuned with respect to pulse sequence. Since our previous expressions were derived for arbitrary $\tau$ and $\phi$, we can easily generalize Eq.~\eqref{phasedifft2t1} to obtain
\begin{align}
&\Phi(t_2,t_1) = \notag \\
&-\sum_m \frac{\gamma b_m}{f_m} [\cos(2\pi f_m t_2 + \phi_m) - \cos(2\phi f_m t_1 + \phi_m)].
\end{align}
Consequently,
\begin{align}
\label{thetaBBmulti}
&\theta_{\text{BB}} = \notag \\
&\sum_m \frac{\gamma b_m}{f_m} \tan(\pi f_m \tau) [\sin \phi_m - \sin(2\pi N f_m \tau + \phi_m)],
\end{align}
and
\begin{align}
\label{thetaCPmulti}
&\theta_{\text{CP}} = \notag \\
&\sum_m \frac{\gamma b_m}{f_m} [\sec(\pi f_m \tau) - 1][\cos(2\pi Nf_m \tau + \phi_m) - \cos \phi_m].
\end{align}
To find $\phi_m$ and $b_m$ is more complicated now. We look at what happens to $\theta_{\text{BB}}$ and $\theta_{\text{CP}}$ as these functions approach any one of $\tau_l = 1/2f_l$ to obtain
\begin{align}
&\theta_{\text{BB}}(\tau\rightarrow 1/2f_l) = \frac{2N\gamma b_l}{f_l}\cos\phi_l \, + \sum_{m\neq l} \frac{\gamma b_m}{f_m} \, \times \notag \\
&\tan(\pi f_m/2f_l) [\sin \phi_m - \sin(\pi N f_m/f_l + \phi_m)], \label{thetaBBtauspecial} \\
&\theta_{\text{CP}}(\tau\rightarrow 1/2f_l) = \frac{2N\gamma b_l}{f_l}\sin\phi_l + \sum_{m\neq l} \frac{\gamma b_m}{f_m}  \, \times \notag \\
&[\sec(\pi f_m/2f_l) - 1] [\cos(\pi N f_m/f_l + \phi_m) - \cos\phi_m] \label{thetaCPtauspecial}. 
\end{align}
Now define the variables $x_1 = b_1 \cos\phi_1$, $x_2 = b_2 \cos\phi_2$, and so on, and $y_1 = b_1 \sin\phi_1$, $y_2 = b_2 \sin\phi_2$, etc. Then the above system of equations becomes a system of linear equations in these new variables. Once we find these new variables, we can find out the values of the different $b_m$ and $\phi_m$ via, for example, $b_1 = \sqrt{x_1^2 + y_1^2}$. Let us illustrate this for a bichromatic magnetic field, namely $B_z(t) = b_1 \sin(2\pi f_1 t + \phi_1) + b_2 \sin(2\pi f_2 t + \phi_2)$. Defining $\boldsymbol{\theta} = [\theta_{\text{BB}}(\tau \rightarrow 1/2f_1), \theta_{\text{BB}}(\tau \rightarrow 1/2f_2), \theta_{\text{CP}}(\tau \rightarrow 1/2f_1), \theta_{\text{CP}}(\tau \rightarrow 1/2f_2)]^T$ and $\mathbf{x} = [b_1 \cos \phi_1, b_2 \cos \phi_2, b_1 \sin \phi_1, b_2 \sin \phi_2]^T$, we find that $\mathbf{x} = \mathbf{A}^{-1}\boldsymbol{\theta}$, where the matrix $\mathbf{A}$ is given by
\begin{align}
A_{11} &= \frac{2N\gamma}{f_1}, \notag \\
A_{12} &= -\frac{\gamma}{f_2} \tan \left(\frac{\pi f_2}{2f_1}\right) \sin\left(\frac{\pi N f_2}{f_1}\right), \notag \\
A_{13} &= 0, \notag \\
A_{14} &= \frac{\gamma}{f_2} \tan \left(\frac{\pi f_2}{2f_1}\right) \left[1 - \cos\left(\frac{\pi N f_2}{f_1}\right)\right], \notag  \\
A_{21} &= -\frac{\gamma}{f_1} \tan \left(\frac{\pi f_1}{2f_2}\right) \sin\left(\frac{\pi N f_1}{f_2}\right), \notag \\
A_{22} &= \frac{2N\gamma}{f_2}, \notag \\
A_{23} &= \frac{\gamma}{f_1} \tan \left(\frac{\pi f_1}{2f_2}\right) \left[1 - \cos\left(\frac{\pi N f_1}{f_2}\right)\right], \notag 
\end{align}
\begin{align}
A_{24} &= 0, \notag \\
A_{31} &= 0, \notag \\
A_{32} &= \frac{\gamma}{f_2} \left[\sec\left(\frac{\pi f_2}{2f_1}\right) - 1 \right]\left[\cos\left(\frac{\pi N f_2}{f_1}\right) - 1 \right], \notag \\
A_{33} &= \frac{2N\gamma}{f_1}, \notag \\
A_{34} &= -\frac{\gamma}{f_2} \left[\sec\left(\frac{\pi f_2}{2f_1}\right) - 1 \right]\sin\left(\frac{\pi N f_2}{f_1}\right), \notag \\
A_{41} &= \frac{\gamma}{f_1} \left[\sec\left(\frac{\pi f_1}{2f_2}\right) - 1 \right]\left[\cos\left(\frac{\pi N f_1}{f_2}\right) - 1 \right], \notag \\
A_{42} &= 0, \notag \\
A_{43} &= -\frac{\gamma}{f_1} \left[\sec\left(\frac{\pi f_1}{2f_2}\right) - 1 \right]\sin\left(\frac{\pi N f_1}{f_2}\right), \notag \\ 
A_{44} &= \frac{2N\gamma}{f_2}.
\end{align}
The off-diagonals of the matrix $\mathbf{A}$ are the contributions of the component of magnetic field that is not tuned with the control pulses.

\begin{figure}[t]
   \includegraphics[scale = 0.6]{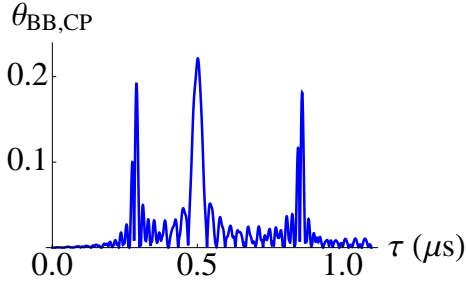}
   \centering
  	\caption{Plot of $\theta_{\text{BB,CP}}$ against $\tau$ for a bichromatic magnetic field. We have used $f_1 = 1 \,\text{MHz}$, $f_2 = 1.75 \, \text{MHz}$, $\phi_1 = \pi/3$, $\phi_2 = \pi/5$, $b_1 = 0.125 \, \mu\text{T}$, and $b_2 = 0.15 \, \mu\text{T}$. We have applied $N = 30$ pulses.}
  	\label{thetaBBCPbichromatic}
\end{figure}

Now suppose that the frequencies $f_1$ and $f_2$ are unknown. This time $\theta_{\text{BB}}$ no longer exhibits a simple periodic behaviour. Instead, we see from Eqs.~\eqref{thetaBBmulti} and \eqref{thetaCPmulti} that $\theta_{BB}$ and $\theta_{\text{CP}}$ should exhibit multiple peaks near each each $\tau_l \rightarrow 1/2f_l$. Therefore, if we plot $\theta_{\text{BB,CP}}$, we can figure out the frequencies by using the positions of the peaks, and the accuracy of our results increases as $N$ increases. Fig.~\ref{thetaBBCPbichromatic} illustrates how this can be done for a bichromatic field. We find three peaks, which are at approximately $\tau = 0.2901 \, \mu$s, $\tau = 0.5016 \, \mu$s and $\tau = 0.8612 \, \mu$s. The first two peaks can be used to calculate the frequencies of bichromatic field as $f_1 \approx 1.724 \, \text{MHz}$ and $f_2 \approx 0.997 \, \text{MHz}$. Note that the third peak (the one near $\tau = 0.86 \, \mu$s) is redundant. This is because, based on our previous considerations, we expect another peak after the first peak with an interval of $1/f_1 = 0.58 \, \mu$s. This is precisely the third peak. 

\section{Effect of decoherence}
\label{sectiondecoherence}

As we have shown, the application of the control pulses leads to a build-up of the phase difference, from which we can then deduce the magnetic field. However, until now, we have ignored the effect of the environment on the NV center. The control pulses also serve as dynamical decoupling pulses, and effectively isolate the NV center from its environment. The decoupling performance of two different pulse sequences in general is not the same. This difference needs to be taken into account if we want to determine the magnetic field accurately.  

Let us now examine the effect of the environment on the NV center closely, following the treatment given in Refs.~\cite{WanganddeLangePRB2012, MkhitaryanarXiv2014}. The decohering spin bath for the NV center is mainly formed by the surrounding nitrogen defects (P1 centers), which are dipolarly coupled to the NV center. This spin bath leads to predominantly pure dephasing, the reason being the large difference between the energies of the NV center and the P1 center. It is difficult to calculate the dynamics of the NV center due to the fact that we have to take into account the complicated dynamics of the P1 centers, which are also dipolarly coupled to each other. In order to make  the problem tractable, a common approximation is to treat the effect of the decohering bath via a classical noise field since the P1 centers are affected negligibly by the NV center. Furthermore, since many P1 centers cause the decoherence of the NV center, the noise field is Gaussian. All in all, we suppose that the NV center experiences a classical Gaussian noise field $B_d(t)$ with zero mean and correlation function 
\begin{equation}
\langle B_d(0) B_d(t) \rangle = \lambda^2 e^{-|t|/\tau_c},
\end{equation}  
where $\tau_c$ is the correlation time and $\lambda$ describes the coupling between the NV center and the P1 centers. 

What is the effect of this noise field? If no pulses are applied, then the transverse spin components decay as 
\begin{equation*}
S(T) = \left\langle \exp \left( -i\int_0^T B_d(t) \, dt \right) \right\rangle.
\end{equation*}
This leads to an exponential decay for large correlation time and weak coupling. The situation changes once control pulses are applied. We now have 
\begin{equation*}
S(T) = \left\langle \exp \left( -i\int_0^T \xi(t) B_d(t) \, dt \right) \right\rangle,
\end{equation*}
where $\xi(t)$, which can assume the values $+1$ or $-1$, takes into account the effect of the pulses by switching sign whenever a pulse is applied. It can then be shown that 
\begin{equation}
S(T) = \exp \left[ -\lambda^2 W(T) \right],
\end{equation}
where $W(T) = \int_0^T e^{-Rs} p(s) \, ds$, with $R = 1/\tau_c$, and $p(s) = \int_0^{T - s} \xi(t) \xi(t + s)\, dt$ depends on the pulse sequence applied. Using this formalism, $W(T)$ can be evaluated for different pulse sequences. For the BB sequence, we can write 
\begin{equation}
W_{\text{BB}}(T) = \Gamma_N(Q_{11}^{\text{BB}} + Q_{12}^{\text{BB}}) - P_N Q_{12}^{\text{BB}},
\end{equation}
with 
\begin{align*}
P_N &= \frac{1 - e^{-N\delta}}{1 - e^{-2\delta}}, \notag \\
\Gamma_N &= \frac{0.5N - (0.5N + 1)e^{-2\delta} + e^{-(N + 2)\delta}}{(1 - e^{-2\delta})^2}, \notag \\
Q_{11}^{\text{BB}} &= \frac{1}{R^2}\left[2\delta - 3 + 4e^{-\delta} - e^{-2\delta} \right], \notag \\
Q_{12}^{\text{BB}} &= \frac{1}{R^2}\left[-1 + 4e^{-\delta} - (2\delta + 3)e^{-2\delta}\right],
\end{align*}
and $\delta = R\tau$. The form of $W_{\text{CP}}(T)$ is similar, with the same $P_N$ and $\Gamma_N$, but we now have 
\begin{align*}
Q_{11}^{\text{CP}} &= \frac{1}{R^2} \left[ 2\delta - 5 + 4(e^{-\frac{\delta}{2}} + e^{-\delta} - e^{-\frac{3\delta}{2}}) + e^{-2\delta}\right], \\
Q_{12}^{\text{CP}} &= \frac{1}{R^2} \left[ 1 - 4(e^{-\frac{\delta}{2}} - e^{-\delta} - e^{-\frac{3\delta}{2}}) - (2\delta + 5)e^{-2\delta}\right].
\end{align*}
With decoherence taken into account, we examine how the previous formalism changes. Equation \eqref{probabilities} gets modified to
\begin{equation}
\label{probwithdecoherence}
p(n|\theta_{\text{BB}}) = \frac{1}{2}[1 + n \sin (\theta_{\text{BB}}) e^{-\lambda^2 W_{\text{BB}}}],
\end{equation}
and an analogous formula exists for the CPMG sequence. Measuring the observable $\sigma_z$, we now obtain $ \langle \sigma_z \rangle = \sin(\theta_{\text{BB}})e^{-\lambda^2 W_{\text{BB}}} \approx \theta_{\text{BB}} e^{-\lambda^2 W_{\text{BB}}}$ for weak magnetic fields. Thus, we have to estimate the magnetic fields using not $\theta_{\text{BB}}$, but rather $\widetilde{\theta}_{\text{BB}} \equiv \sin(\theta_{\text{BB}}) e^{-\lambda^2 W_{\text{BB}}} \approx \theta_{\text{BB}}e^{-\lambda^2 W_{\text{BB}}}$. We can then write
\begin{align}
\widetilde{\theta}_{\text{BB}} = &\frac{\gamma b}{f} \tan (\pi f \tau) \, \times \notag \\
&[\sin \phi - \sin(2\pi N f \tau + \phi)] e^{-\lambda^2 W_{\text{BB}}}, \\
\widetilde{\theta}_{\text{CP}} = &\frac{\gamma b}{f} [\sec(\pi f \tau) - 1] \, \times \notag \\
&[\cos(2\pi N f \tau + \phi) - \cos \phi] e^{-\lambda^2 W_{\text{CP}}}.
\end{align}
It then follows that
\begin{align}
\widetilde{\theta}_{\text{BB}}(\tau \rightarrow 1/2f) &= \frac{2N\gamma b}{f} \cos \phi \, e^{-\lambda^2 W_{\text{BB}}(\tau \rightarrow 1/2f)}, \\
\widetilde{\theta}_{\text{CP}}(\tau \rightarrow 1/2f) &= \frac{2N\gamma b}{f} \sin \phi \, e^{-\lambda^2 W_{\text{CP}}(\tau \rightarrow 1/2f)},
\end{align}
leading to
\begin{align}
\tan \phi &= \frac{\widetilde{\theta}_{\text{CP}} e^{\lambda^2 W_{\text{CP}}}}{\widetilde{\theta}_{\text{BB}} e^{\lambda^2 W_{\text{BB}}}},\\
b &= \frac{f}{2N\gamma} \sqrt{[\widetilde{\theta}_{\text{CP}} e^{\lambda^2 W_{\text{CP}}}]^2 + [\widetilde{\theta}_{\text{BB}} e^{\lambda^2 W_{\text{BB}}}]^2},
\end{align}
where $\widetilde{\theta}_{\text{BB}}$, $\widetilde{\theta}_{\text{CP}}$, $W_{\text{BB}}$ and $W_{\text{CP}}$ are calculated for $\tau \rightarrow 1/2f$. Thus our previous results on finding the amplitude and phase should be adjusted by taking into account decoherence effects, which can be done since the values of $\tau_c$ and $\lambda$ can be obtained experimentally \cite{WanganddeLangePRB2012}. 

The adjustment for multichromatic fields is carried out in a similar manner. Equations \eqref{thetaBBtauspecial} and \eqref{thetaCPtauspecial} are modified to
\begin{align}
&\widetilde{\theta}_{\text{BB}}(\tau\rightarrow 1/2f_l) = e^{-\lambda^2 W_{\text{BB}}}\bigg\lbrace\frac{2N\gamma b_l}{f_l}\cos\phi_l \, + \sum_{m\neq l} \frac{\gamma b_m}{f_m} \, \times \notag \\
&\tan(\pi f_m/2f_l) [\sin \phi_m - \sin(\pi N f_m/f_l + \phi_m)]\bigg\rbrace,\\
&\widetilde{\theta}_{\text{CP}}(\tau\rightarrow 1/2f_l) = e^{-\lambda^2 W_{\text{CP}}}\bigg\lbrace\frac{2N\gamma b_l}{f_l}\sin\phi_l + \sum_{m\neq l} \frac{\gamma b_m}{f_m}  \, \times \notag \\
&[\sec(\pi f_m/2f_l) - 1] [\cos(\pi N f_m/f_l + \phi_m) - \cos\phi_m]\bigg\rbrace,
\end{align}
where $W_{\text{BB}}$ and $W_{\text{CP}}$ are evaluated at the corresponding $\tau \rightarrow 1/2f_l$. The matrix $\mathbf{A}$ gets modified accordingly.

\begin{figure}[t]
   \includegraphics[scale = 0.5]{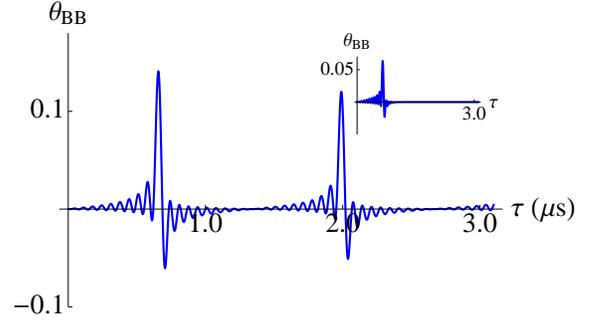}
   \centering
  	\caption{Plot of $I_{bb}^{\text{BB}}$ against $\tau$. We have used $f = 0.75 \,\text{MHz}$, $\phi = \pi/5$, $b = 0.1 \, \mu\text{T}$, $\tau_c = 25 \, \mu$s, $\lambda = 0.36 \, \mu\text{s}^{-1}$ and $N = 20$. For the inset, the same parameters are used except that now $\lambda = 3.6\, \mu\text{s}^{-1}$. For $\lambda = 3.6\, \mu\text{s}^{-1}$ and $\tau_c = 25 \, \mu$s, the dephasing time (without pulses) is $T_2 \approx 2.8 \, \mu$s \cite{deLangePRL2011,WanganddeLangePRB2012}, while with $\lambda = 0.36\, \mu\text{s}^{-1}$ and $\tau_c = 25 \, \mu$s, we have $T_2 \approx 13.2 \, \mu$s. Even longer dephasing times have been obtained experimentally \cite{RondinarXiv2013}.}
  	\label{figthetaBBwithdecoherence}
\end{figure} 

\begin{figure}[b]
   \includegraphics[scale = 0.5]{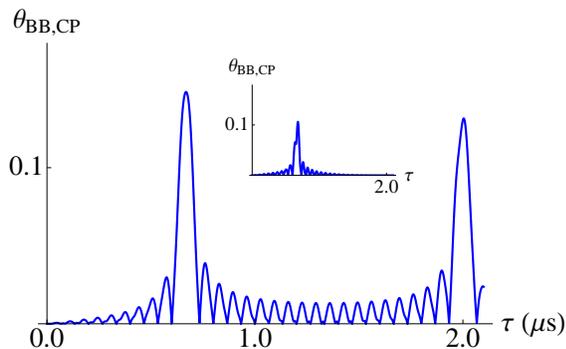}
   \centering
  	\caption{Plot of $I_{bb}^{\text{BB,CP}}$ against $\tau$. The parameters used are the same as Fig.~\ref{figthetaBBwithdecoherence}. In particular, for the main figure $\lambda = 0.36\, \mu\text{s}^{-1}$, while for the inset $\lambda = 3.6\, \mu\text{s}^{-1}$.}
  	\label{figthetawithdecoherenceBBCP}
\end{figure} 

What about figuring out the frequency? We have checked numerically that for relatively weak decoherence (that is, for NV centers with relatively long dephasing times), our previously proposed methods work. For example, in Fig.~\ref{figthetaBBwithdecoherence}, the first peak is located at $\tau \approx 0.6567\, \mu$s, while the second peak is at $\tau \approx 1.9900 \, \mu$s, which gives us $750$ kHz as an extremely accurate estimate of the frequency. On the other hand, for shorter dephasing time [see inset of Fig.~\ref{figthetaBBwithdecoherence}], the first proposed method to find the frequency for a monochromatic field, namely finding the interval between two peaks for $\theta_{\text{BB}}$, fails - decoherence causes the second peak to be negligible. However, the second method still allows us to obtain reasonable estimates of the frequency for both weak and strong decoherence as illustrated in Fig.~\ref{figthetawithdecoherenceBBCP}. For weak decoherence [see the main figure], the first peak is located at $\tau \approx 0.6687 \, \mu$s leading to a frequency estimate of $f \approx 748$ kHz. For stronger decoherence [see inset], we obtain the frequency as $735$ kHz. Of course, even better estimates can be obtained if $\tau_c$ and $\lambda$ are known.

\subsection{Fisher information analysis with decoherence}

We now investigate the Fisher information matrix in the presence of decoherence. Due to decoherence, the Fisher information matrix elements do not keep on increasing as $N$ is increased. Rather, there is now a competition between the effects of decoherence and the increased sensing time. This can be shown by deriving the Fisher information matrix using Eq.~\eqref{probwithdecoherence}. Carrying out the calculations as before, we find that 
\begin{equation}
\label{Fisherwithdecoherence}
[I^{\text{BB}}(\mathbf{y})]_{mn} = \frac{\partial\theta_{\text{BB}}}{\partial y_m} \frac{\partial\theta_{\text{BB}}}{\partial y_n} e^{-2\lambda^2 W_{\text{BB}}}.
\end{equation}
The same result is obtained for the quantum Fisher information. Thus there is an exponential suppression factor that makes the Fisher information negligible for large $N$. This should be compared with the case without decoherence, where the Fisher information keeps on increasing as $N$ is increased.

 Suppose now that we only estimating the magnetic field amplitude $b$ for a monochromatic magnetic field using the BB sequence (we are assuming that the phase and the frequency are known). The number of pulses $N$ that should be applied is then chosen such that the Fisher information $I_{bb}^{\text{BB}}$ is maximized. We know that, for $\tau \rightarrow 1/2f$,
\begin{equation*}
I_{bb}^{\text{BB}} = \frac{4N^2 \gamma^2}{f^2} \cos^2\phi \, e^{-2\lambda^2 W_{\text{BB}}}.
\end{equation*}

\begin{figure}[t]
   \includegraphics[scale = 0.6]{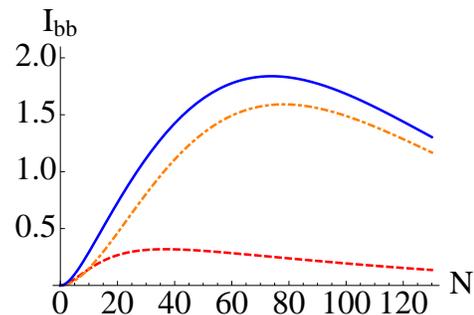}
   \centering
  	\caption{Plot of $I_{bb}^{\text{BB}}$ (dashed, red), $I_{bb}^{\text{CP}}$ (dot-dashed, orange) and $I_{bb}^{\text{BB,CP}}$ (solid, blue) against $N$. We have used $f = 0.75 \,\text{MHz}$, $\phi = \pi/5$, $b = 0.1 \, \mu\text{T}$, $\lambda = 3.6 \, \mu\text{s}^{-1}$ and $\tau_c = 25\,\mu$s.}
  	\label{figIBBCP}
\end{figure}

\begin{figure}[t]
   \includegraphics[scale = 0.6]{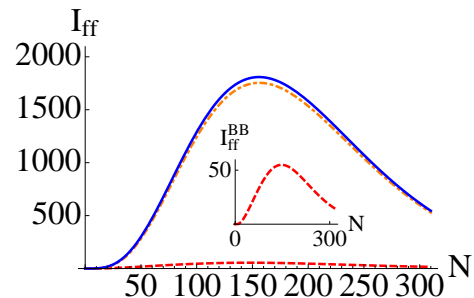}
   \centering
  	\caption{Plot of $I_{ff}^{\text{BB}}$ (dashed, red), $I_{ff}^{\text{CP}}$ (dot-dashed, orange) and $I_{ff}^{\text{BB,CP}}$ (solid, blue) against $N$. The inset shows the detailed behaviour of $I_{ff}^{\text{BB}}$. The parameters used are the same as Fig.~\ref{figIBBCP}.}
  	\label{figIffBBCP}
\end{figure} 

Behavior of $I_{bb}^{\text{BB}}$ as $N$ changes is illustrated in Fig.~\ref{figIBBCP}. We see that as $N$ initially increases, the Fisher information increases due to the increase in sensing time. However, after a certain number of pulses, $N_0^{\text{BB}}$, it starts to decrease due to decoherence effects kicking in. To derive an expression for $N_0^{\text{BB}}$, we use the fact that 
\begin{align*}
\frac{\partial P_N}{\partial N} &= \frac{\delta}{1 - e^{-2\delta}} e^{-N\delta}, \\
\frac{\partial \Gamma_N}{\partial N} &= \frac{0.5 - 0.5e^{-2\delta} - \delta e^{-2\delta} e^{-N\delta}}{(1 - e^{-2\delta})^2}, 
\end{align*}
and that $\delta \ll 1$ experimentally to find that $N_0^{\text{BB}}$ can be found by numerically solving the equation
\begin{equation}
96\tau_c f^3 - N_0^{\text{BB}}\lambda^2 (1 + 3 e^{-N_0^{\text{BB}}/2f\tau_c}) = 0.
\end{equation}
This leads to $N_0^{\text{BB}} = 36$ for the parameters used in Fig.~\ref{figIBBCP}.

In a similar manner, the optimal number of pulses to be used if the CP scheme is used is found to be 
\begin{equation}
N_0^{\text{CP}} = \frac{96 \tau_c f^3}{\lambda^2},
\end{equation}
which gives $N_0^{\text{CP}} = 78$ for the parameters used in Fig.~\ref{figIBBCP}. Note that $N_0^{\text{CP}} > N_0^{\text{BB}}$ because the CPMG scheme is able to suppress decoherence more effectively. Thus, the optimal number of pulses to be used if both pulse sequences are used to deduce the amplitude of the magnetic field $b$ is somewhere in between $N_0^{\text{BB}}$ and $N_0^{\text{CP}}$ (see the solid, blue line in Fig.~\ref{figIBBCP} - the optimal value when both the BB and CPMG sequences are used is $N_0^{\text{BB,CP}} = 74$). The analysis for $I_{\phi \phi}$ is exactly analogous. For $I_{ff}$, on the other hand, taking $\tau$ to be close to $1/2f$, we find that an increase in sensing time is now more beneficial. Thus, as shown in Fig.~\ref{figIffBBCP}, the number of pulses that should be used is greater than before. In particular, we find that now $N_0^{\text{BB}} = 148$ and $N_0^{\text{CP}} = 156$. Furthermore, once again we see that the better performance of the CPMG sequence in suppressing decoherence leads to much higher values of the Fisher information.

\begin{figure}[t]
   \includegraphics[scale = 0.6]{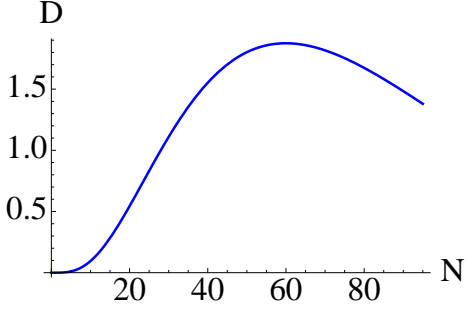}
   \centering
  	\caption{Plot of $D$ against $N$. Here we have used $f = 0.75 \,\text{MHz}$, $\phi = \pi/3$, $b = 1.0 \, \mu\text{T}$, $\lambda = 3.6 \, \mu\text{s}^{-1}$ and $\tau_c = 25\,\mu$s.}
  	\label{figDmonochromatic}
\end{figure}

Until now, we have been optimizing $N$ so as to increase the precision of estimating a single parameter. However, as we have shown, we can estimate $b$ and $\phi$ if we use both BB and CPMG sequences. It is important to note that now $I^{\text{BB,CP}}_{b\phi} \neq 0$ due to the different performance of the two pulse sequences in suppressing decoherence. What is the best $N$ to use in this case? This question is more complicated to answer because the Fisher information is no longer a simple scalar. In such a case, the usual approach is to minimize some real-valued function of the Fisher information matrix. For example, one option, known as D-optimality, is to maximize the determinant of the Fisher information matrix \cite{Atkinsonbook}. Doing so minimizes the volume of the uncertainty ellipsoid described by the inverse of the Fisher information matrix. An example is shown in Fig.~\ref{figDmonochromatic}, where we have plotted $D \equiv \text{det}(\mathbf{I}^{\text{BB,CP}})$ as a function of $N$. We see that according to this criterion, we should use approximately $N = 60$ in order to obtain the best estimates of the parameters $b$ and $\phi$. 

\subsection{Fisher information for multichromatic fields}

\begin{figure}[b]
   \includegraphics[scale = 0.6]{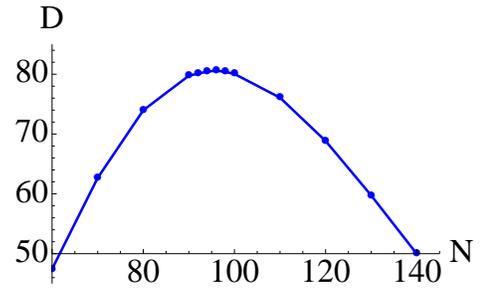}
   \centering
  	\caption{Plot of $D$ against $N$. The parameters used are $\lambda = 3.6\,\mu\text{s}^{-1}$, $\tau_c = 25\,\mu s$, $f_1 = 0.75\, \text{MHz}$, $f_2 = 1\,\text{MHz}$, $b_1 = 1.0\,\mu$T, $b_2 = 1.5\,\mu$T, $\phi_1 = \pi/5$ and $\phi_2 = \pi/3$.}
  	\label{figDbichromatic}
\end{figure} 

We now extend the analysis of the Fisher information matrix to multichromatic fields. As we have shown, if the $M$ frequency components are known, then the field can be estimated using the $M$ values $\widetilde{\theta}_{\text{BB}}(\tau \rightarrow 1/2f_i)$ and the $M$ values $\widetilde{\theta}_{\text{CP}}(\tau \rightarrow 1/2f_i)$. Suppose that we want to estimate $b_1,\hdots,b_M$ and $\phi_1,\hdots,\phi_M$. In order to obtain the best estimates of the these parameters, we once again need to calculate the Fisher information matrix, which is now of dimension $2M \times 2M$. The total Fisher information matrix can be calculated by first calculating the Fisher information for each pulse sequence, and then adding up these $2M$ matrices. Each of these individual matrices can be calculated using the fact that 
\begin{align*}
&\frac{\partial\theta_{\text{BB}}}{\partial b_k} (\tau \rightarrow 1/2f_l) = \frac{2N\gamma}{f_k}\cos\phi_k \delta_{kl} \, + \\ \notag 
&(1 - \delta_{kl})\frac{\gamma}{f_k}\tan\left(\frac{\pi f_k}{2f_l}\right) \left[\sin\phi_k - \sin\left(\frac{\pi Nf_k}{f_l} + \phi_k \right)\right], \\
&\frac{\partial\theta_{\text{BB}}}{\partial \phi_k} (\tau \rightarrow 1/2f_l) = -\frac{2N\gamma b_k}{f_k}\sin\phi_k \delta_{kl} \, + \\ \notag 
&(1 - \delta_{kl})\frac{\gamma b_k}{f_k}\tan\left(\frac{\pi f_k}{2f_l}\right) \left[\cos\phi_k - \cos\left(\frac{\pi Nf_k}{f_l} + \phi_k \right)\right], \\
&\frac{\partial\theta_{\text{CP}}}{\partial b_k} (\tau \rightarrow 1/2f_l) = \frac{2N\gamma}{f_k}\sin\phi_k \delta_{kl} \, + (1 - \delta_{kl}) \, \times \\ \notag 
&\frac{\gamma}{f_k}\left[\sec\left(\frac{\pi f_k}{2f_l}\right) - 1\right] \left[\cos\left(\frac{\pi Nf_k}{f_l} + \phi_k \right) - \cos \phi_k\right], \\
&\frac{\partial\theta_{\text{CP}}}{\partial \phi_k} (\tau \rightarrow 1/2f_l) = \frac{2N\gamma b_k}{f_k}\cos\phi_k \delta_{kl} \, + (1 - \delta_{kl}) \, \times \\ \notag 
&\frac{\gamma b_k}{f_k}\left[\sec\left(\frac{\pi f_k}{2f_l}\right) - 1\right] \left[\sin \phi_k - \sin\left(\frac{\pi Nf_k}{f_l} + \phi_k \right)\right],
\end{align*}
and thereafter taking decoherence into account [see Eq.~\ref{Fisherwithdecoherence}]. Once again, the components of the magnetic field that are not tuned with the control pulses play a prominent role. The inverse of the total Fisher information matrix then gives us bounds on the variances of the parameters that are being estimated, which we then optimize with respect to the number of pulses that need to be applied. We carried out this process for a bichromatic field, once again using D-optimality, and the results are illustrated in Fig.~\ref{figDbichromatic}. We see that $N = 96$ is the best choice for estimating the parameters $b_1$, $b_2$, $\phi_1$ and $\phi_2$.

Finally, before concluding, it should be noted that pulse errors can become important when the number of pulses becomes large. In this case, it is better to change the pulse sequences that we are applying from single-axis control to two-axis control \cite{WanganddeLangePRB2012}. What this means is that instead of implementing $\pi$ pulses as only $e^{-i\pi \sigma_x/2}$, we use alternately use $e^{-i\pi \sigma_x/2}$ and $e^{-i\pi \sigma_y/2}$. Such two-axis control is known to appreciably reduce pulse errors.

\section{Conclusion}

In this paper, we have proposed experimentally implementable methods employing simple pulse sequences applied to NV centers in order to determine the amplitude, phase and frequency of unknown weak magnetic fields. We started by deriving expressions for the phase difference developed by a NV center in the presence of a monochromatic field with pulses applied according to the BB scheme and the CPMG scheme. In particular, our expressions are valid for arbitrary pulse spacing and magnetic field phase. We then showed how these expressions could be used to determine the amplitude, phase and frequency of monochromatic magnetic fields. Since our expressions take into account the possibility that the pulses applied are not tuned to the frequency of the magnetic field, they were generalized in a straightforward manner for multichromatic magnetic fields. We then showed that for multichromatic magnetic fields with $M$ frequency components, $2M$ measurements can be used to determine the magnetic field if the frequencies are known. We also discussed how to determine the frequency of the fields with excellent accuracy. 

Throughout, we also calculated the Fisher information matrix to show how the sensitivity of the estimation of the parameters improved as the number of pulses applied increased (or, in other words, the total sensing time increased). However, in reality, the NV center is interacting with its surrounding environment of nitrogen defects and carbon nuclei. This interaction leads to decoherence, which means that the superposition state of the NV center is eventually lost. By treating the environment as a classical noise field, we showed how decoherence implies that the sensitivity does not keep on increasing as the number of pulses increases. Rather, one needs to obtain the optimal number of pulses that should be applied such that the effect of increased sensing time and the influence of decoherence are balanced. By using parameters from recent experiments, we calculated the optimal number of pulses that should be used for the estimation of various parameters. We also generalized these results to multichromatic fields, and discussed how the optimization can be carried out in this case, with an explicit example give for a bichromatic magnetic field. It is hoped that these results are useful in the determination of weak magnetic fields, and in particular weak multichromatic magnetic fields, with fewer resources required.

\begin{acknowledgements}
This work is supported by the Singapore National Research Foundation under NRF Grant No. NRF-NRFF2011-07. Discussions with M.~Tsang and R.~Nair are gratefully acknowledged. 
\end{acknowledgements}

\appendix

\section{Proof of Eq.~\eqref{firstidentity}}

We want to show that 
\begin{align*}
S = &\sum_{k = 1}^{N - 1} (-1)^k \cos (2\pi k f \tau + \phi)  = -\frac{1}{2} \sec (\pi f \tau) \, \times \notag \\&[\cos(\pi f \tau + \phi) + \cos(2\pi N f \tau - \pi f \tau + \phi)].
\end{align*}
We start by noting that 
\begin{align*}
S = -&\sum_{k = 0}^{N/2 - 1} \cos[2\pi f\tau(2k + 1) + \phi] \, + \\
&\sum_{k = 0}^{N/2 - 2} \cos[2\pi f\tau(2k + 2) + \phi].
\end{align*}
Now let $S' = \sum_{k = 0}^{N/2 - 1} \cos[2\pi f \tau(2k + 1) + \phi]$. Then, using Lagrange's trigonometric identities,
\begin{equation*}
S' = \frac{\sin(\pi N f \tau)}{\sin(2\pi f \tau)} \cos(\pi N f\tau + \phi).
\end{equation*}
Similarly,
\begin{align*}
&\sum_{k = 0}^{N/2 - 2} \cos[2\pi f \tau(2k + 2) + \phi] = \\
&\frac{\sin(N\pi f \tau - 2\pi f \tau)}{\sin(2\pi f\tau)}\cos(N\pi f \tau + \phi),
\end{align*}
from which it then follows that 
\begin{align*}
S &= \frac{\cos(N\pi f \tau + \phi)}{\sin(2\pi f \tau)} [\sin(N\pi f \tau - 2\pi f \tau) - \sin(N\pi f \tau)] \\ 
&= -\sec(\pi f \tau) \cos(N\pi f \tau - \pi f \tau) \cos(N\pi f \tau + \phi). 
\end{align*}
Thus,
\begin{align*}
S= &-\frac{1}{2}\sec(\pi f \tau) \times \notag \\
&[\cos(\pi f \tau + \phi) + \cos(2\pi N f \tau - \pi f \tau + \phi)].
\end{align*}
The proof of Eq.~\eqref{secondidentity} is very similar.

\end{document}